\documentclass[aps,prl,twocolumn,showpacs,groupedaddress]{revtex4}

\usepackage[english]{babel}
\usepackage{amsfonts,bm,amsmath,amssymb,mathrsfs}

\newcommand{\eq}{\begin{equation}}
\newcommand{\eeq}{\end{equation}}

\begin{document}

\title{Perturbed Kerr Black Holes can probe deviations from General Relativity}

\author{Enrico Barausse$^{1}$ and Thomas P. Sotiriou$^{2}$}
\affiliation{$^1$ SISSA, International School for
             Advanced Studies, and INFN, Via Beirut 2, 34014 Trieste, Italy\\
$^2$ Center for Fundamental Physics, University of Maryland, College Park, MD 20742-4111, USA}

\date{\today}
\begin{abstract}
Although the Kerr solution is common to many gravity theories, 
its perturbations are different in different theories. Hence, perturbed Kerr black holes can probe deviations from General Relativity. 
\end{abstract}

\maketitle

It has been claimed~\cite{psaltis} that the existence of the Kerr(-de Sitter) solution in many modified gravity theories implies that its verification around an astrophysical black hole (BH) cannot probe deviations from General Relativity (GR). This would have implications for current attempts to test GR  using BHs \cite{psaltis}. 
While this is true for electromagnetic observations, we argue that gravitational wave (GW) experiments can
in principle distinguish Kerr BHs in GR and in modified gravity theories.

Ref.~\cite{psaltis} considered $f(R)$ gravity, quadratic gravity and vector-tensor gravity. 
These theories are built so as to have maximally symmetric solutions (Minkowski or de Sitter) in vacuum, and this is done by imposing that the field equations reduce to GR if $R_{,\mu}=0$, $R_{\mu\nu}=R\,g_{\mu\nu}/4$ and $T_{\mu\nu}=0$. As such, 
it had long been known that Kerr-de Sitter, which satisfies these conditions, is a solution in these theories.
Moreover, these theories admit vacuum solutions different from Kerr, as noted also in Ref.~\cite{psaltis}.
[The only exception is Palatini $f(R)$ gravity, which reduces to GR in vacuum.] 
Astrophysical BHs form by gravitational collapse, and there is yet no guarantee that this 
will lead to a Kerr BH if this solution is not unique. 

However, we will set this (potentially important) issue aside, as done in Ref.~\cite{psaltis},  
and show that even a Kerr BH permits probing deviations from GR.
Indeed, the (vacuum) metric perturbations over a Kerr BH in the theories considered in Ref.~\cite{psaltis} behave differently from GR. Again, the exception is Palatini $f(R)$ gravity, as this reduces to GR in vacuum. 
We use metric $f(R)$ gravity \textit{as an example}, and assume
$f(0)\!=\!0$: this is required to have a Kerr (and \textit{not} Kerr-de Sitter) BH~\cite{psaltis}, which has $R\!=\!R_{\mu\nu}\!=\!0$
thus simplifying the calculation. With $\Box\!\equiv\!\nabla^\nu\nabla_\nu$ and $'\!\equiv\! \partial/\partial R$,  the vacuum field equations are
\eq\label{eq:fR_field}
f'(R)R_{\mu\nu}-\frac{1}{2}f(R) g_{\mu\nu}-(\nabla_\mu\nabla_\nu -g_{\mu\nu} \Box) f'(R) =0.
\eeq
We denote the metric perturbation by $h_{\mu\nu}$ and
use the \textit{Lorenz gauge}, defined by
$\nabla_\nu \bar{h}^{\mu\nu}=0$, 
with $\bar{h}_{\mu\nu}\equiv h_{\mu\nu}-h g_{\mu\nu}/2$.
This is always possible: $\bar{h}^{\mu\nu}$ transforms as
\eq\label{eq:gauge}
\hat{\bar{h}}_{\mu\nu}=\bar{h}_{\mu\nu}-\nabla_\mu \xi_\nu-\nabla_\nu \xi_\mu+g_{\mu\nu} \nabla_\alpha \xi^\alpha\,,
\eeq
and we only need to impose
$\Box\xi^\mu=\nabla_\nu \bar{h}^{\mu\nu}$ \cite{scott_lectures}. 
Perturbing eq.~(\ref{eq:fR_field}) over a Kerr background in this gauge, we have
\eq\label{eq:pert_eq}
\Box \bar{h}_{\mu\nu}+2 R_{\mu\alpha\nu\beta} \bar{h}^{\alpha\beta}=
-\lambda\left(\nabla_\mu\nabla_\nu-g_{\mu\nu}\Box\right)\Box\bar{h}\,,
\eeq
where $\lambda\equiv{f''(0)}/{f'(0)}$.
In  GR $f'(0)=1$ and $f''(0)=0$, and thus $\lambda=0$. 
Also, in GR it is possible to set $\bar{h}=0$, although only in a globally vacuum spacetime, using the residual freedom of the Lorenz gauge
(one can perform a transformation with $\xi^{\mu}$
satisfying $\Box\xi^\mu=0$)~\cite{scott_lectures}.
From eq.~(\ref{eq:gauge}) it follows that
we need $\nabla_\mu \xi^\mu+\bar{h}/2=0$ in order to have $\hat{\bar{h}}=0$. Taking the ``box'' of $\nabla_\mu \xi^\mu+\bar{h}/2$, and using
the trace of eq.~\eqref{eq:pert_eq} and $\Box\xi^\mu\!=\!R\!=\!R_{\mu\nu}\!=\!0$,
one can show that $\nabla_\mu \xi^\mu+\bar{h}/2$ satisfies the homogeneous wave equation only if $f''(0)=0$ (as in GR): it is then possible to choose initial data for $\xi^\mu$ on a Cauchy hypersurface such that $\nabla_\mu \xi^\mu+\bar{h}/2$ and its derivative normal to the hypersurface vanish, thus ensuring $\nabla_\mu \xi^\mu+\bar{h}/2=0$ everywhere. This is not possible if $f''(0)\neq0$, and $\bar{h}$ cannot then be zeroed (even in globally vacuum spacetimes). Thus, eq.~(\ref{eq:pert_eq}) differs from its GR analog. For example, 
over a Minkowski background, besides the
propagation modes of GR, eq.~(\ref{eq:pert_eq}) also has a plane-wave solution 
$\bar{h}_{\mu\nu}\!\propto\! (\eta_{\mu\nu}+k_\mu k_\mu/m^2) \exp(i k_\alpha x^\alpha)$,
where $\boldsymbol{k}\equiv \omega(\kappa)\, \partial_t+ \kappa\,\boldsymbol{n}$ ($\boldsymbol{n}$ being  
the propagation direction) and $\omega(\kappa)^2\equiv \kappa^2+m^2$ (with $m^2\!\equiv\!(3 \lambda)^{-1}\!>\!0$: if 
$\lambda\!<\!0$ the gravity theory is non-viable~\cite{dolgov}). 
These GWs cannot be zeroed in the Lorenz gauge, correspond to
massive gravitons with velocity $d\omega/d\kappa\!<\!1$ (detectable by LISA~\cite{LISA}) and their polarization differs from GR. Note that this mode corresponds to scalar field excitations [{\em cf.} equivalent Brans-Dicke theory with a potential~\cite{bd}]. It is also present in 
Brans-Dicke theory with no potential \cite{BD} and it affects the orbital evolution of binary systems~\cite{LISA,BD}. 

In conclusion: while the Kerr solution is common to many gravity theories, 
its perturbations are not. 
Because GWs from perturbed Kerr BHs behave differently in different theories,
their detection can be used to discriminate them, and the concerns of
Ref.~\cite{psaltis} seem unjustified.

We acknowledge discussions with L. Barack, J. Gair, T. Jacobson, J. C. Miller, D. Psaltis and L. Rezzolla.

\end{document}